\documentclass[10pt,oneside,final]{IEEEtran}
\usepackage{amssymb}
\usepackage{amsmath}

\usepackage{graphicx}
\usepackage{color}
\usepackage{multicol}

\usepackage{subfigure}
\usepackage{bm}
\usepackage{latexsym}
\usepackage{stfloats}
\usepackage{float}
\usepackage{exscale}
\usepackage{relsize}
\usepackage{cite}
\usepackage{cases}
\usepackage{algorithm}
\usepackage{algpseudocode}
\usepackage{amsthm}
\usepackage{epsfig}
\usepackage{epstopdf}
\usepackage{breqn}
\usepackage{enumerate}
\usepackage{multirow}
\usepackage[utf8]{inputenc}
\usepackage{algorithmicx}
\usepackage{amssymb}
\begin{document}
	\title{ Energy Minimization for Active RIS-Aided UAV-Enabled SWIPT Systems }
	\author{Zhangjie Peng,
		Ruijing Liu,
		Cunhua Pan,~\IEEEmembership{Member,~IEEE},
		Zhenkun Zhang, and 
		Jiangzhou Wang,~\IEEEmembership{Fellow,~IEEE}\vspace{-0.5cm}
		\thanks{Z. Peng and R. Liu are with the College of Information, Mechanical and Electrical Engineering, Shanghai Normal University, Shanghai 200234, China. (e-mail: pengzhangjie@shnu.edu.cn; 1000511821@smail.shnu.edu.cn).}
		\thanks{C. Pan and Z. Zhang are with the National Mobile Communications Research Laboratory, Southeast University, Nanjing 210096, China. (e-mail: cpan@seu.edu.cn; zhenkun\underline{~}zhang@seu.edu.cn).}
		\thanks{J. Wang is with the School of Engineering, University of Kent, CT2 7NT Canterbury, U.K. (e-mail: j.z.wang@kent.ac.uk).}

	}
	\maketitle
	\newtheorem{lemma}{Lemma}
	\newtheorem{theorem}{Theorem}
	\newtheorem{remark}{Remark}
	\newtheorem{corollary}{Corollary}
	\newtheorem{proposition}{Proposition}\vspace{-0.1cm}
	\begin{abstract} 
		 In this paper, we consider an active reconfigurable intelligent surface (RIS)-aided unmanned aerial vehicle (UAV)-enabled simultaneous wireless information and power transfer (SWIPT)  system with multiple ground users. Compared with the conventional passive RIS, the active RIS deploying the internally integrated amplifiers can offset part of the multiplicative fading. In this system, we deal with an optimization problem of minimizing the total energy cost of the UAV. Specifically, we alternately optimize the trajectories, the hovering time, and the reflection vectors at the active RIS by using the successive convex approximation (SCA) method. Simulation results show that the active RIS performs better in energy saving than the conventional passive RIS.
	\end{abstract}

	\section{Introduction}
	As one of the potential technologies driving the development of 6G, the reconfigurable intelligent surface (RIS), which consists of a large number of reflecting elements, has become a promising technology to enhance the communication quality of wireless networks. With the characteristics of low cost and easy deployment, researchers have studied the deployment of the RIS into wireless communication systems \cite{1,2,4}.
	
	Due to the high flexibility, unmanned aerial vehicle (UAV) has been proposed to assist the signal transmission in the communication system  \cite{6,7}. 
Meanwhile, simultaneous wireless information and power transfer (SWIPT), which can enhance the energy and information transmission, has emerged as a promising communication transmission approach for internet-of-things (IoT) networks \cite{10}. By integrating the RIS, UAV and SWIPT, the performance of the communication system can be further enhanced.
In \cite{9}, an RIS was deployed in SWIPT systems to enhance both information and energy transmission.
The authors in \cite{11} considered a problem for maximizing the average achievable rate, and showed that the performance of SWIPT systems with a single IoT device can be improved with the deployment of both the UAV and RIS.
	
	However, the performance of the RIS-aided systems is affected by the significant multiplicative fading on the cascaded channels of the reflective links. To solve this problem, the active RIS integrates the amplifiers into the reflecting elements to offset the multiplicative fading \cite{12}. The authors in \cite{13} theoretically compared the active RIS-aided system with the passive RIS-aided system, and demonstrated that the better performance can be achieved by the active RIS-aided system when the power budget is adequate. In \cite{8}, the authors investigated an active RIS-aided system and proposed a joint computing and communication design using the successive convex approximation (SCA) method.
	
	Against the above background, we consider an active RIS-aided UAV-enabled SWIPT system with multiple ground users, where the UAV serves as a transmitter. In this paper, we aim to minimize the total energy cost of the UAV while fulfilling the energy reception and information transmission target for all users. The problem is solved by reformulating the original problem into two subproblems, and applying the SCA method to handle the subproblems, and the original problem is solved by alternately optimizing two subproblems. Simulation results show that the active RIS performs better in terms of energy saving than the passive RIS under the same total energy supply for the UAV.
	
		\begin{figure}[t]
		\centering
		\includegraphics[width=0.9\linewidth]{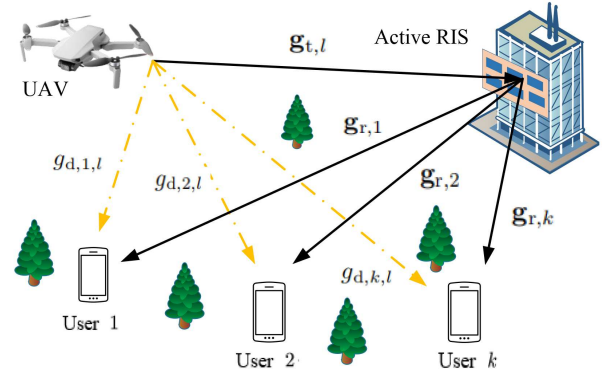}\vspace{-0.3cm}
		\caption{{Active RIS-aided UAV-enabled SWIPT system.} } 
		\label{fig1}\vspace{-0.6cm}
	\end{figure}

	\section{System Model And Problem Formulation}
	\subsection{System Model}
	As shown in Fig. \ref{fig1}, we consider an active RIS-aided UAV-enabled SWIPT system with ${K}$ ground users, where the active RIS consists of ${M}$ reflecting elements, and both the UAV and each user are equipped with one antenna. To facilitate the representation of distances between devices, we use a complex number ${q=x+jy}$ to denote the  horizontal coordinate $(x,y)$ of the device. 
	In this case, the position of the active RIS and user ${k}$ are respectively denoted by ${q_{\rm R}=q_{\rm R}^x+jq_{\rm R}^y}$ with the height ${H_{\rm R}}$ and ${q_{{\rm S},k}=q_{{\rm S},k}^x + jq_{{\rm S},k}^y}$ for ${ k \in  \mathcal{K} \triangleq \left\{ 1,...,K \right\}}$ with zero height. 
	Meanwhile, we set the UAV flying at the same height, and each flight consists of ${L}$ straight path segments starting from ${q_{{\rm V}, 0}}$ and ending at ${q_{{\rm V}, L}}$.
	Besides, the hovering position of the UAV is denoted by ${q_{{\rm V},l}=q^x_{{\rm V},l} + jq^y_{{\rm V},l}}$ for ${l \in \mathcal{L} \triangleq\left\lbrace 1,...,L-1\right\rbrace }$ with the height ${H_{\rm V}}$. 
	Then, the trajectory of the UAV is expressed as ${\mathbf{q}_{\rm V} \!\!=\left[ q_{{\rm V}, 0},..., q_{{\rm V}, l},..., q_{{\rm V}, L}\right] ^{\rm T} \!\!\!\in \mathbb{C}^{\left( L+1\right) \times 1}}$, and the hovering time is denoted by ${\mathbf{t} = [t_1, t_l, ..., t_{L-1}]^{\rm T}}$.
	
	When the UAV is hovering at the ${l}$-th hovering position, the reflection vector and the diagonal reflection matrix are respectively denoted by ${\pmb{\phi}_l \!\!= \!\!\left[b_{1, l}e^{j{\theta}_{1, l}}, ..., b_{M, l}e^{j{\theta}_{M, l}}\right] ^{\rm T}}$ and ${\mathbf{\Theta}_l \! ={\rm diag} \left( \pmb{\phi}_l \right)=\!{\rm diag}\!\left( b_{1, l}e^{j{\theta}_{1, l}}\!, ..., b_{M, l}e^{j{\theta}_{M, l}}\right)}$, where ${{\theta}_{m, l}}$ and ${b_{m,l}}$ are the phase shift and the amplitude of the ${m}$-th reflecting element at the active RIS, respectively.
	The channels from the UAV to user ${k}$, from the UAV to the active RIS, and from the active RIS to user ${k}$ are respectively denoted by ${g_{{\rm d}, k, l} \!\in \!\mathbb{C}^{1 \!\times \!1}}$, ${\mathbf{g}_{{\rm t}, l} \!\in \!\mathbb{C}^{M \!\times \!1}}$ and ${\mathbf{g}_{{\rm r}, k} \!\in \!\mathbb{C}^{M \!\times \!1}}$. The Rician fading channels ${g_{{\rm d}, k, l}}$, ${\mathbf{g}_{{\rm t}, l}}$ and ${\mathbf{g}_{{\rm r}, k}}$ are respectively modeled as
	\begin{align}
	\label{gdkl}g_{{\rm d}, k, l}=&\sqrt{\beta_{{\rm d}, k, l}}\left( \sqrt{\frac{\mu_{\rm d}}{\mu_{\rm d}+1}}g_{{\rm d}, k, l}^{\rm LoS}+\sqrt{\frac{1}{\mu_{\rm d}+1}}g_{{\rm d}, k, l}^{\rm NLoS}\right), \\
	\label{gtl}\mathbf{g}_{{\rm t}, l}\!\!\!\!\!\quad=&\sqrt{\beta_{{\rm t}, l}}\!\quad\left( \sqrt{\frac{\mu_{\rm t}}{\mu_{\rm t}+1}}\mathbf{g}_{{\rm t}, l}^{\rm LoS}\!\!\!\!\quad+\sqrt{\frac{1}{\mu_{\rm t}+1}}\mathbf{g}_{{\rm t}, l}^{\rm NLoS}\right), \\
	\label{grk}\mathbf{g}_{{\rm r}, k}=&\sqrt{\beta_{{\rm r}, k}}\!\!\quad\left( \sqrt{\frac{\mu_{\rm r}}{\mu_{\rm r}+1}}\mathbf{g}_{{\rm r}, k}^{\rm LoS}\!\!\!\!\quad+\sqrt{\frac{1}{\mu_{\rm r}+1}}\mathbf{g}_{{\rm r}, k}^{\rm NLoS}\right), 
	\end{align}
	where each element of the non-line-of-sight (NLoS) components ${g_{{\rm d}, k, l}^{\rm NLoS}}$, ${\mathbf{g}_{{\rm t}, l}^{\rm NLoS}}$ and ${\mathbf{g}_{{\rm r}, k}^{\rm NLoS}}$ are i.i.d. circularly symmetric complex Gaussian distribution with zero mean and unit variance,  ${\mu_{\rm d}}$, ${\mu_{\rm t}}$ and ${\mu_{\rm r}}$ are the Rician factors, ${g_{{\rm d}, k, l}^{\rm LoS}}$, ${\mathbf{g}_{{\rm t}, l}^{\rm LoS}}$ and ${\mathbf{g}_{{\rm r}, k}^{\rm LoS}}$ are the line-of-sight (LoS) components. Under the  fly-hover-broadcast (FHB) protocol \cite{7}, it is assumed that the UAV only transmits signals during hovering mode, and thus the large-scale fading coefficients ${\beta_{{\rm d}, k, l}}$, ${\beta_{{\rm t}, l}}$ and ${\beta_{{\rm r}, k}}$ that depend on the hovering position are respectively given by
	\begin{align}
		\beta_{{\rm d}, k, l} &=\frac{\beta_0}{(\left\vert q_{{\rm V},l} \!-\! q_{{\rm S}, k} \right\vert^2 \!\!+\!\! H_{\rm V}^2) ^{\tau_{\rm d}/2}}\triangleq\frac{\beta_0}{d_{{\rm d}, k, l}^{\tau_{\rm d}}},\\
		\beta_{{\rm t}, l}&=\frac{\beta_0}{( \left\vert q_{{\rm V},l}-q_{\rm R} \right\vert^2+\left( H_{\rm V}-H_{\rm R}\right) ^2) ^{\tau_{\rm t}/2}}\triangleq\frac{\beta_0}{d_{{\rm t}, l}^{\tau_{\rm t}}},\\
		\beta_{{\rm r}, k}&=\frac{\beta_0}{( \left\vert q_{{\rm S}, k}-q_{\rm R} \right\vert^2+H_{\rm R}^2) ^{\tau_{\rm r}/2}}\triangleq\frac{\beta_0}{d_{{\rm r}, k}^{\tau_{\rm r}}},
	\end{align}
	where ${\beta_0}$ is the channel gain of ${1\,\rm m}$, ${\tau_{\rm d}}$, ${\tau_{\rm t}}$ and ${\tau_{\rm r}}$ are the path-loss coefficients, ${d_{{\rm d}, k, l}}$, ${d_{{\rm t}, l}}$ and ${d_{{\rm r}, k}}$ are the distances from user ${k}$ to the UAV, from the UAV to the active RIS and from the active RIS to user ${k}$, respectively. ${g_{{\rm d}, k, l}^{\rm LoS}}$, ${\mathbf{g}_{{\rm t}, l}^{\rm LoS}}$ and ${\mathbf{g}_{{\rm r}, k}^{\rm LoS}}$ represent the LoS components under the uniform linear array (ULA) model, which are respectively expressed as
	\begin{align}
	\!\!\! g_{{\rm d}, k, l}^{\rm LoS}&=e^{-j\frac{2\pi}{\lambda}d_{{\rm d}, k, l}},\\
		\!\!\!\mathbf{g}_{{\rm t}, l}^{\rm LoS}&=\! e^{-j\frac{2{\pi}d_{{\rm t}, l}}{\lambda}}\!\left[\!1,\! e^{-j\frac{2{\pi}d}{\lambda}\cos{\omega_{{\rm t}, l}}},\!...,\! e^{-j\frac{2{\pi}\left( M-1\right) d}{\lambda}\cos{\omega_{{\rm t}, l}}}\!\right]^{\rm T}\!\!\!\!,\\
		\!\!\!\!\mathbf{g}_{{\rm r}, k}^{\rm LoS}&=\! e^{-j\frac{2{\pi}d_{{\rm r}, k}}{\lambda}}\!\!\left [\!1,\! e^{-j\frac{2{\pi}d}{\lambda}\cos{\omega_{{\rm r}, k}}}\!,\!...,\! e^{-j\frac{2{\pi}\left( M-1\right) d}{\lambda}\cos{\omega_{{\rm r}, k}}}\!\right]^{\rm T}\!\!\!\!,
	\end{align}
	where ${\lambda}$ and ${d}$ respectively represent the wavelength and the element spacing of the active RIS, ${\cos{\omega_{{\rm t}, l}}=\frac{q_{\rm R}^x-q_{{\rm V},l}^x}{d_{{\rm t}, l}}}$ and ${\cos{\omega_{{\rm r}, k}}=\frac{q_{{\rm S}, k}^x-q_{\rm R}^x}{d_{{\rm r}, k}}}$ are the cosine of angle-of-arrival (AoA) and angle-of-departure (AoD), respectively. 
	
	In our system, the information transfer and energy supplication can be obtained simultaneously, and the power split ratio for decoding information is ${\eta \in [0,1]}$ and for harvested power is  ${1-\eta }$.
	Then, the received signal of user ${k}$ is given by
	\begin{align}\label{signal}
		y_k=&\sqrt{p_k}(\mathbf{g}_{{\rm r}, k}^{\rm H}\mathbf{\Theta}_l \mathbf{g}_{{\rm t}, l}\!+\!g_{{\rm d}, k, l})x_k+\mathbf{g}_{{\rm r}, k}^{\rm H}\mathbf{\Theta}_l\mathbf{n}_{\rm RIS}+n_{\rm r}\notag\\
		&+\!\!\!\!\!\!\sum\limits_{j\ne k, j \in \mathcal{K}}\!\!\!\sqrt{p_j}(\mathbf{g}_{{\rm r}, k}^{\rm H}\mathbf{\Theta}_l \mathbf{g}_{{\rm t}, l}\!+\!g_{{\rm d}, k, l})x_j\notag\\
		=&\sqrt{p_k}g_{k, l}x_k+\!\!\!\!\!\!\sum\limits_{j\ne k, j \in \mathcal{K}}\!\!\!\sqrt{p_j}g_{k, l}x_j+\mathbf{g}_{{\rm r}, k}^{\rm H}\mathbf{\Theta}_l\mathbf{n}_{\rm RIS}+n_{\rm r},
	\end{align}
	where ${g_{k, l}\!=\!\mathbf{g}_{{\rm r}, k}^{\rm H}\mathbf{\Theta}_l \mathbf{g}_{{\rm t}, l}\!+\!g_{{\rm d}, k, l}}$, ${p_k}$ is the transmit power for user ${k}$ from the UAV, ${x_k\!\sim\!\mathcal{C}\mathcal{N}\!\left( 0, 1\right)}$ is the transmit signal for user ${k}$, ${n_{\rm r}\sim\mathcal{C}\mathcal{N}\left( 0, \delta_{\rm r}^2\right) }$ is the noise at user $k$, and ${\mathbf{n}_{\rm RIS}\sim\mathcal{C}\mathcal{N}\left( 0, \delta_{\rm RIS} ^2\mathbf{I}\right)}$ is the noise at the active RIS.
	When the UAV is hovering at the ${l}$-th hovering position, the ergodic rate of user ${k}$ is given by
	\begin{align}
	\!\!\! R_{k,l} \!=\! \mathbb{E}\!\left\{\!\!{\rm log}_2\!\!\left(\!\!1\!\!+\!\frac{\eta p_k | g_{k, l} |^2 }{\!\!\!\!\sum\limits_{j\ne k, j \in \mathcal{K}}\!\!\!\!\!\eta p_j | g_{k, l} |^2+\delta_{\rm RIS} ^2\| \mathbf{g}_{{\rm r}, k}^{\rm H}\mathbf{\Theta}_l \|^2+\delta_{\rm r}^2}\!\!\right) \!\!\!\right\}\!.\!
	\end{align}
 By utilizing the derivations in \cite[Lemma 1]{15}, ${{R}_{k,l}}$ of user ${k}$ can be approximated as
\begin{align}
	R_{k,l} \approx \tilde{R}_{k,l}={\rm log}_2\left(1+{\rm SINR}_{k,l}\right) ,
\end{align}
where  ${{\rm SINR}_{k,l}}$ is given by
\begin{align}
	\!\!\!\!\!\!{\rm SINR}_{k,l} \!=\!\frac{\eta p_k\mathbb{E}\{ | g_{k, l} |^2 \}}{\!\!\!\!\!\!\!\sum\limits_{j\ne k, j \in \mathcal{K}}\!\!\!\!\!\!\!\eta p_j\mathbb{E}\{ | g_{k, l} |^2\}\!\!+\!\delta_{\rm RIS} ^2\mathbb{E}\{\| \mathbf{g}_{{\rm r}, k}^{\rm H}\mathbf{\Theta}_l \|^2\}\!\!+\!\delta_{\rm r}^2}.\!\!\!\!\!
\end{align}

    Furthermore, the harvested power ${\tilde{P}_{k,l}}$ at user ${k}$ is given by
	\begin{align}\label{pkl}
		\tilde{P}_{k,l}=(1-\eta)p_k\mathbb{E}\{\left\vert g_{k, l} \right\vert^2\}.
	\end{align}

	 When the UAV flies at a speed of ${v}$, the propulsion power is modeled as
	\begin{align}
	 \!\!\! P_{\rm p}(v)\!=\! P_0 \!\!\left( \!\!1 \!+\!\frac{3v^2}{U_{\rm tip}^2}\right) \!\! + \!\! P_{\rm i}\!\left( \!\!\sqrt{1 \!\!+\!\!\frac{v^4}{4v_0^4}}\!-\!\frac{v^2}{2v_0^2}\right) ^{\frac{1}{2}}\!\!\!\!+\!\frac{1}{2}d_0{\rho}sAv^3\!\!,\!\!\!
	\end{align}
	where the constants such as ${P_0}$, ${P_{\rm i}}$ and ${v_0}$ represent the configuration of the UAV, and their definitions are detailed in \cite{7}. 
	
	Then, the total energy cost of the UAV can be given by
	\begin{align}
		\!\!\!\!\!\!\! E_{\rm V}\!\left( \mathbf{q}_{\rm V} \! , \mathbf{t}\right) \!=\! P_{\rm p}(v)\!\sum\limits_{l\in \mathcal{L}} \!\frac{\left\vert q_{{\rm V}, l}\!-\! q_{{\rm V}, l-1} \right\vert}{v}\!+\!\!\! \sum\limits_{l\in \mathcal{L}}\!\!\left( P_{\rm p}^{\rm hov}\!\!+ \! P_{\rm t} \right)\! t_l,\!\!
	\end{align}
	where ${P_{\rm p}^{\rm hov} = P_{\rm p}(0) = P_0 + P_{\rm i}}$ is the hovering power, and ${P_t}$ is the radiated power of the UAV. 
\vspace{-1ex}
	\subsection{Problem Formulation }
	\vspace{-1ex}
  We consider a problem of minimizing the total energy cost of the UAV, subject to the SINR threshold ${\gamma_k}$ at user ${k}$, the energy requirement ${E_k^{\rm req}}$ at user ${k}$, and the energy limitation ${E_{\rm RIS}^{\rm act}}$ at the active RIS. Then, the problem is formulated as
	\begin{subequations}\label{swipt_Problem1}
		\begin{align}
			\min_{{\mathbf{q}_{\rm V}}, {\{\pmb{\phi}_l\}},\mathbf{t}} \label{objective function1} &\quad  E_{\rm V}\left( \mathbf{q}_{\rm V}, \mathbf{t}\right)
			\\
			\mathrm{s}.\mathrm{t}.\!\quad\label{swipt_Problem1-b}&\quad {\rm SINR}_{k,l}\geqslant \gamma_k,\, k\in\mathcal{K},\, l\in {\mathcal{L}},
			\\
			\label{swipt_Problem1-c} &\quad \sum_{l\in {\mathcal{L}}}{ t_l\tilde{P}_{k,l}}\geqslant E_{k}^{\rm req} ,\, k\in\mathcal{K},
			\\
			\label{swipt_Problem1-d}&\quad \mathbb{E} \{ \left\| \mathbf{\Theta}_l \mathbf{s}_l \right\|^2 \}t_l \leqslant E_{\rm RIS}^{\rm act},\, l\in {\mathcal{L}},
		\end{align}
	\end{subequations}
	In \eqref{swipt_Problem1-d}, ${\mathbf{s}_l \!=\!\sum_{k \in \mathcal{K}} \!\!\sqrt{p_k}\mathbf{g}_{{\rm t},l}x_k \!+\!\mathbf{n}_{\rm RIS}}$ and ${\mathbf{\Theta}_l \! ={\rm diag} \left( \pmb{\phi}_l \right)}$.

%\begin{proof}
%	See Appendix \ref{Theorem1}.
%\end{proof}

%\emph{Proof:}
%See Appendix \ref{Theorem1}.{\hfill $\blacksquare$\par}

	\section{Joint Reflection Coefficients And Trajectory Design}
In this section, we propose an effective SCA-based algorithm for solving Problem \eqref{swipt_Problem1}. Specifically, we reformulate the original problem into two subproblems, and solve the subproblems by using SCA method. Then, the original problem is solved by alternately optimizing two subproblems. 
\vspace{-2ex}
	\subsection{The Design of Reflection Coefficient Vectors}
   In this subsection, we first optimize ${\left\{\pmb{\phi}_l\right\}}$ with given ${\mathbf{q}_{\rm V}}$ and ${\mathbf{t}}$ when the UAV is at the ${l}$-th hovering position. Since \eqref{objective function1} is not a function of ${\left\{\pmb{\phi}_l\right\}}$, the corresponding subproblem is a feasibility-check problem. To improve the convergence performance, the optimization objective can be enhanced by maximizing the total oversupplied energy of all users \cite{1}. Since the subproblem may not be feasible, it may cause the algorithm to converge early. To solve this, we set the objective function to maximize the minimum charged energy of all users instead \cite{1}, and the subproblem is formulated as 
   \begin{subequations}\label{SubProblem1}
   	\begin{align}
   		\max_{{\{\pmb{\phi}_l\}},\varepsilon} &\quad  \varepsilon
   		\\
   		\mathrm{s}.\mathrm{t}.\label{energy4} &\quad (1/E_{k}^{\rm req})\sum_{l\in {\mathcal{L}}}{ t_l\tilde{P}_{k,l}}\geqslant \varepsilon,\, k\in\mathcal{K},
   		\\
   		&\quad \eqref{swipt_Problem1-b},\,\eqref{swipt_Problem1-d}, \notag
   	\end{align}
   \end{subequations}
   where ${\varepsilon}$ is an auxiliary variable. 
   
   However, Problem \eqref{SubProblem1} is still non-convex due to the non-convex constraints \eqref{swipt_Problem1-b} and \eqref{energy4}.
   In the following, we formulate constraints \eqref{energy4} and \eqref{swipt_Problem1-d} into the more tractable forms, and then derive constraints \eqref{swipt_Problem1-b} and \eqref{energy4} into the convex constraints.
   
   Firstly, we need to reformulate constraint \eqref{energy4} into a more tractable form. By substituting \eqref{gdkl}, \eqref{gtl} and \eqref{grk} into the term ${\mathbb{E}\{\left\vert g_{k, l} \right\vert^2\}}$ of ${\tilde{P}_{k,l}}$ in \eqref{pkl}, we have
   \begin{align}
   	\!\!\!\!\!\mathbb{E} \left\{ \left| \left( \mathbf{g}_{{\rm r},k}^{\rm H}\mathbf{\Theta}_l  \mathbf{g}_{{\rm t},l}\!+\! {g}_{{\rm d},k,l} \right) \right|^2 \right\} \!\!&=\!\!\mathbb{E}\!\left\{ \left| a_{0,k,l} \right|^2 \right\} \!\!+\!\mathbb{E}\! \left\{ \left| a_{1,k,l} \right|^2 \right\} \notag\\
   	+\mathbb{E} \left\{ \left| a_{2,k,l} \right|^2 \right\} \!\!&+\!\mathbb{E}\! \left\{ \left| a_{3,k,l} \right|^2 \right\} \!\!+\!\mathbb{E} \!\left\{ \left| a_{4,k,l} \right|^2 \right\}\!\!, 
   \end{align}
   where
   \begin{align}
    &\!\!\! a_{0,k,l}\!=\!\!\sqrt{\!\!\frac{\mu _{\rm r}\beta _{{\rm r},k}}{\mu _{\rm r}\!+\!1}}\!\sqrt{\!\!\frac{\mu _{\rm t}\beta _{{\rm t},l}}{\mu _{\rm t}\!+\!1}}\!\!\left(\! \mathbf{g}_{{\rm r},k}^{\rm LoS} \right) ^{\rm H}\!\!\mathbf{\Theta}_l \mathbf{g}_{{\rm t},l}^{\rm LoS}\!\!+\!\!\sqrt{\!\!\frac{\mu _{\rm d}\beta _{{\rm d},k,l}}{\mu _{\rm d}+1}}{g}_{{\rm d},k,l}^{\rm LoS}, \!\!\\
   	&\!\!\!a_{1,k,l}\!=\!\!\sqrt{\frac{\beta _{{\rm d},k,l}}{\mu _{\rm d}+1}}{g}_{{\rm d},k,l}^{\rm NLoS},\\
   	&\!\!\!a_{2,k,l}\!=\!\!\sqrt{\frac{\beta _{{\rm r},k}}{\mu _{\rm r}+1}}\sqrt{\frac{\mu _{\rm t}\beta _{{\rm t},q}}{\mu _{\rm t}+1}}\left( \mathbf{g}_{{\rm r},k}^{\rm NLoS} \right) ^{\rm H} \!\!\mathbf{\Theta}_l \mathbf{g}_{{\rm t},l}^{\rm LoS}, \\
   	&\!\!\!a_{3,k,l}\!=\!\!\sqrt{\frac{\mu _{\rm r}\beta _{{\rm r},k}}{\mu _{\rm r}+1}}\sqrt{\frac{\beta _{{\rm t},l}}{\mu _{\rm t}+1}}\left( \mathbf{g}_{{\rm r},k}^{\rm LoS} \right) ^H \mathbf{\Theta}_l \mathbf{g}_{{\rm t},l}^{\rm NLoS}, \\
   	&\!\!\!a_{4,k,l}\!=\!\!\sqrt{\frac{\beta _{{\rm r},k}}{\mu _{\rm r}+1}}\sqrt{\frac{\beta _{{\rm t},l}}{\mu _{\rm t}+1}}\left( \mathbf{g}_{{\rm r},k}^{\rm NLoS} \right) ^{\rm H}\!\!\mathbf{\Theta}_l \mathbf{g}_{{\rm t},l}^{\rm NLoS}. 
   \end{align}
   \begin{figure*}[hb]
   	   	\hrulefill
   	\begin{align}\label{a0}
   		a_{0,k,l}=&\sqrt{\frac{\mu _{\rm r}\beta _{{\rm r},k}}{\mu _{\rm r}+1}}\sqrt{\frac{\mu _{\rm t}\beta _{{\rm t},l}}{\mu _{\rm t}+1}}\left( e^{-j\frac{2\pi d_{{\rm r},k}}{\lambda}}\left[ 1,e^{-j\frac{2\pi d}{\lambda}\cos \omega _{{\rm r},k}},\cdots ,e^{-j\frac{2\pi \left( M-1 \right) d}{\lambda}\cos \omega _{{\rm r},k}} \right] ^{\rm T} \right) ^{\rm H}\cdot {\rm diag}(b_1,...,b_M)\notag\\
   		&\cdot {\rm diag}(e^{j\theta _{1,l}},...,e^{j\theta _{M,l}}) \cdot
   		e^{-j\frac{2\pi d_{{\rm t},l}}{\lambda}}\!\!\left[ 1,e^{-j\frac{2\pi d}{\lambda}\cos \omega _{{\rm t},l}},\cdots ,e^{-j\frac{2\pi \left( M-1 \right) d}{\lambda}\cos \omega _{{\rm t},l}} \right] ^{\rm T} \!\!\!+\!\sqrt{\frac{\mu _{\rm d}\beta _{{\rm d},k,l}}{\mu _{\rm d}+1}}e^{-j\frac{2\pi d_{{\rm d},k,l}}{\lambda}} \notag
   		\\
   		=&\sqrt{\frac{\mu _{\rm r}\mu _{\rm t}\beta _{{\rm r},k}\beta _{{\rm t},l}}{\left( \mu _{\rm r}+1 \right) \left( \mu _{\rm t}+1 \right)}}\sum_{m=1}^M{b_me^{j( \frac{2\pi ( d_{{\rm r},k}-d_{{\rm t},l} ) +2\pi ( m-1 ) d( \cos \omega _{{\rm r},k}-\cos \omega _{{\rm t},l} )}{\lambda}+\theta _{m,l} )}} \!\!\!\quad+\sqrt{\frac{\mu _{\rm d}\beta _{{\rm d},k,l}}{\mu _{\rm d}+1}}e^{-j\frac{2\pi d_{{\rm d},k,l}}{\lambda}} \notag
   		\\
   		=&e^{-j\frac{2\pi d_{{\rm d},k,l}}{\lambda}}\left( \sqrt{\frac{\mu _{\rm d}\beta _{{\rm d},k,l}}{\mu _{\rm d}+1}}+\sqrt{\frac{\mu _{\rm r}\mu _{\rm t}\beta _{{\rm r},k}\beta _{{\rm t},l}}{\left( \mu _{\rm r}+1 \right) \left( \mu _{\rm t}+1 \right)}}\sum_{m=1}^M{b_me^{j\left( \psi _{k,m,l}+\theta _{m,l} \right)}} \right), 
   	\end{align}
   	\hrulefill
   	\begin{align}\label{Ea0}
   		\!\!\mathbb{E}\{\left| a_{0,k,l} \right|^2\}&=\frac{\mu _{\rm d}\beta _{{\rm d},k,l}}{\mu _{\rm d}+1}\!+\!\frac{\mu _{\rm r}\mu _{\rm t}\beta _{{\rm r},k}\beta _{{\rm t},l}}{\left( \mu _{\rm r} \!+\!1 \right) \left( \mu _{\rm t} \! +\!1 \right)}\left| \sum_{m=1}^M{b_me^{j\left( \psi _{k,m,l}+\theta _{m,l} \right)}} \right|^2\!\!\!+\!2\mathrm{Re}\left\{ \sqrt{\frac{\mu _{\rm r}\mu _{\rm t}\beta _{{\rm r},k}\beta _{{\rm t},l}}{\left( \mu _{\rm r}+1 \right) \left( \mu _{\rm t}+1 \right)}}\sum_{m=1}^M{b_me^{j\left( \psi _{k,m,l}+\theta _{m,l} \right)}} \right\} \notag
   		\\
   		&=\frac{\mu _{\rm r}\mu _{\rm t}\beta _{{\rm r},k}\beta _{{\rm t},l}}{\left( \mu _{\rm r}+1 \right) \left( \mu _{\rm t}+1 \right)}\pmb{\phi} _{l}^{\rm H}\pmb{\psi} _{k,l}\pmb{\psi} _{k,l}^{\rm H}\pmb{\phi} _l+2\sqrt{\frac{\mu _{\rm r}\mu _{\rm t}\beta _{{\rm r},k}\beta _{{\rm t},l}}{\left( \mu _{\rm r}+1 \right) \left( \mu _{\rm t}+1 \right)}}\mathrm{Re}\left\{ \pmb{\psi} _{k,l}^{\rm H}\pmb{\phi} _l \right\} +\frac{\mu _{\rm d}\beta _{{\rm d},k,l}}{\mu _{\rm d}+1}. 
   	\end{align}
   \end{figure*}
   As ${\mathbb{E} \{ \mathbf{g}_{{\rm r},k}^{\rm NLoS}(\mathbf{g}_{{\rm r},k}^{\rm NLoS} ) ^{\rm H} \} =\mathbf{I}_M}$, we have ${\mathbb{E}\{\left| a_{1,k,l} \right|^2\}=\frac{\beta _{{\rm d},k,l}}{\mu _{\rm d}+1}}$, ${\mathbb{E}\{\left| a_{2,k,l} \right|^2\}=\frac{\mu _{\rm t}\beta _{{\rm r},k}\beta _{{\rm t},l}}{\left( \mu _{\rm r}+1 \right) \left( \mu _{\rm t}+1 \right)}\pmb{\phi} _{l}^{\rm H}\pmb{\phi} _l}$, ${\mathbb{E}\{\!\left| a_{3,k,l} \!\right|^2\!\}\!\!=\!\!\frac{\mu _{\rm r}\beta _{{\rm r},k}\beta _{{\rm t},l}}{\left(\! \mu _{\rm r}\!+1\! \right) \left( \mu _{\rm t}\!+1\! \right)}\pmb{\phi} _{l}^{\rm H}\!\pmb{\phi} _l}$, and ${\mathbb{E}\{\!\left| a_{4,k,l}\! \right|^2\!\}\!\!=\!\!\frac{\beta _{{\rm r},k}\beta _{{\rm t},l}}{\left(\! \mu _{\rm r}\!+1\! \right) \left( \!\mu _{\rm t}\!+1\! \right)}\pmb{\phi} _{l}^{\rm H}\!\pmb{\phi} _l}$. The expansion of term ${a_{0,k,l}}$ and the derivation of ${\mathbb{E}\{\left| a_{0,k,l} \right|^2\}}$ are respectively shown in \eqref{a0} and \eqref{Ea0}, and ${\pmb{\psi} _{k,l}}$ in \eqref{Ea0} is given by
      \begin{align}  
   	\pmb{\psi} _{k,l}\triangleq &\left[ \begin{matrix}
   		e^{-j\psi _{k,1,l}},&		\cdots&		,\\
   	\end{matrix}e^{-j\psi _{k,M,l}} \right]^{\rm T},
   \end{align}
   where
\begin{align}
   	\!\!\!\!\!\psi _{k,m,l}\triangleq &\frac{1}{\lambda}(2\pi \left( d_{{\rm d},k,l}+d_{{\rm r},k}-d_{{\rm t},l} \right) \notag\\
   	&+2\pi \left( m-1 \right) d\left( \cos \omega _{{\rm r},k}-\cos \omega _{{\rm t},l} \right)).
   \end{align}
   
    Then, ${\tilde{P}_{k,l}}$ in \eqref{pkl} can be reformulated as
   	\begin{align}\label{energy1}
   	\tilde{P}_{k,l}=\left( 1-\eta \right) p_k\mathcal{D}_{k,l}, 
   \end{align}
where
\begin{align}
	   	\mathcal{D}_{k,l} \!\triangleq &  \mathbb{E}\{\left\vert g_{k, l} \right\vert^2\} =\pmb{\phi} _{l}^{\rm H}\mathbf{A}_{k,l}\pmb{\phi} _l+2\mathrm{Re}\left\{ \mathbf{a}_{k,l}^{\rm H}\pmb{\phi} _l \right\} +\beta _{{\rm d},k,l},\\
	   	\!\!\!\!\mathbf{A}_{k,l}\!\triangleq &\frac{\mu _{\rm r}\mu _{\rm t}\beta _{{\rm r},k}\beta _{{\rm t},l}}{\left( \mu _{\rm r}\!\!+\!1 \right) \!\left( \mu _{\rm t} \!\!+\!1 \right)}\pmb{\psi} _{k,l}\pmb{\psi} _{k,l}^{\rm H}\!\!+\!{\frac{\left( \mu _{\rm r}\!\!+\!\!\mu _{\rm t}\!\!+\!\!1 \right)\! \beta _{{\rm r},k}\beta _{{\rm t},l}}{\left( \mu _{\rm r}\!+\!1 \right) \!\left( \mu _{\rm t}\!+\!1 \right)}}\!\mathbf{I}_M, \\
	   	\mathbf{a}_{k,l}\triangleq &\sqrt{\frac{\mu _{\rm d}\mu _{\rm r}\mu _t\beta _{{\rm d},k,l}\beta _{{\rm r},k}\beta _{{\rm t},l}}{\left( \mu _{\rm d}+1 \right) \left( \mu _{\rm r}+1 \right) \left( \mu _{\rm t}+1 \right)}}\pmb{\psi} _{k,l}.
\end{align}

   Therefore, by defining  ${h_k(\pmb{\phi})\!\triangleq \!(1/E_{k}^{\rm req})\!\sum_{l\in \mathcal{L}}{\! t_l\tilde{P}_{k,l}}}$, constraint \eqref{energy4} can be reformulated as the following tractable form.
   \begin{align}\label{SubProblem1-1-b}
   	h_k(\pmb{\phi}) \geqslant \varepsilon,\, k\in\mathcal{K}.
   \end{align}
where
\begin{align}
	\!\!\! h_k \! \left( \pmb{\phi} \right) \!=&\pmb{\phi} ^{\rm H}\hat{\mathbf{A}}_k\pmb{\phi} +2\mathrm{Re}\left\{ \hat{\mathbf{a}}_{k}^{\rm H}\pmb{\phi} \right\} +{\zeta}_{1, k},\\
	\pmb{\phi}\!\!\!\! \quad\triangleq& \left[ \pmb{\phi} _{1}^{\rm T},\cdots ,\pmb{\phi} _{L-1}^{\rm T} \right] ^{\rm T},\\
	{\zeta}_{1, k}\triangleq&\sum_{l\in {\mathcal{L}}}{\frac{(1-\eta) p_kt_l}{E_{k}^{\rm req}}\beta _{{\rm d},k,l}},\\
	\!\!\!\!\hat{\mathbf{A}}_k\triangleq & {\rm diag} \left(\!\!\frac{\left( 1\!-\!\eta \right)\! p_kt_1A_{k,1}}{E_{k}^{\rm req}},\!...,\!\frac{\left( 1\!-\!\eta \right)\! p_kt_{L-1}A_{k,L-1}}{E_{k}^{\rm req}}\!\!\right)\!\!,\!\! \\
	\hat{\mathbf{a}}_k\triangleq&\left[ \frac{\left( 1- \!\eta \right) p_kt_1}{E_{k}^{\rm req}}a_{k,1}^{\rm T},\cdots ,\!\frac{\left( 1-\eta \right) p_kt_{L-1}}{E_{k}^{\rm req}}a_{k,L-1}^{\rm T} \right] ^{\rm T} \!\!\!\!.
\end{align}

As the term ${\mathbb{E} \{ \left\| \mathbf{\Theta}_l \mathbf{s}_l \right\|^2 \}}$ in \eqref{swipt_Problem1-d} is difficult to handle, we reformulate it  into a more tractable form as 
\begin{align}
	&\mathbb{E} \left\{ \left\| \mathbf{\Theta}_l\left( \sum_{k\in\mathcal{K}}\sqrt{p_k}\mathbf{g}_{{\rm t},l}x_k+\mathbf{n}_{\rm RIS} \right) \right\|^2 \right\}\notag
	\\
	&=\Big(\sum_{k\in\mathcal{K}}{p_k}\mathbb{E} \Big\{ \!\!\sqrt{\beta _{{\rm t},l}}\left( \sqrt{\frac{\mu _{\rm t}}{\mu _{\rm t}+1}}\mathbf{g}_{{\rm t},l}^{\rm LoS}+\sqrt{\frac{1}{\mu _{\rm t}+1}}\mathbf{g}_{{\rm t},l}^{\rm NLoS} \right) ^{\rm H} \notag
	\\
	&\!\!\!\!\quad\mathbf{\Theta}_l ^{\rm H} \!\mathbf{\Theta}_l \sqrt{\!\beta _{{\rm t},l}}\left(\! \sqrt{\!\frac{\mu _{\rm t}}{\mu _{\rm t} \!+\!1}}\mathbf{g}_{{\rm t},l}^{\rm LoS}\!\!+\!\sqrt{\!\frac{1}{\mu _{\rm t} \!+\!1}}\mathbf{g}_{{\rm t},l}^{\rm N \! LoS} \right)\!\! \Big\} \!+\!\delta _{\rm RIS}^{2}\pmb{\phi} _{l}^{\rm H}\pmb{\phi} _l \!\!\Big)\! \notag
	\\
	&=\left(\sum_{k\in\mathcal{K}}{p_k}\left( \frac{\beta _{{\rm t},l}\mu _{\rm t}}{\mu _{\rm t}+1}\pmb{\phi} _{l}^{\rm H}\pmb{\phi} _l+\frac{\beta _{{\rm t},l}}{\mu _{\rm t}+1}\pmb{\phi} _{l}^{\rm H}\pmb{\phi} _l \right) +\delta _{\rm RIS}^{2}\pmb{\phi} _{l}^{\rm H}\pmb{\phi} _l\right) \notag
	\\
	\label{Egtheta}&=\left( \sum_{k\in\mathcal{K}}{p_k}\beta _{{\rm t},l}+\delta _{\rm RIS}^{2} \right) \pmb{\phi} _{l}^{\rm H}\pmb{\phi} _l.
\end{align}
Then, constraint \eqref{swipt_Problem1-d} can be reformulated as
\begin{align}\label{ris-energy}
  \left( \sum_{k\in\mathcal{K}}{p_k}\beta _{{\rm t},l}+\delta _{\rm RIS}^{2} \right) \pmb{\phi} _{l}^{\rm H}\pmb{\phi} _lt_l\leqslant E^{\rm act}_{\rm RIS},\, l\in {\mathcal{L}}.
\end{align}

Note that constraints \eqref{swipt_Problem1-b} and \eqref{SubProblem1-1-b} are still non-convex. 
By using the similar processing procedure in \eqref{Egtheta}, we can obtain ${\mathbb{E}\{\| \mathbf{g}_{{\rm r}, k}^{\rm H}\mathbf{\Theta}_l \|^2\} = \beta _{{\rm r},k}\pmb{\phi} _{l}^{\rm H}\pmb{\phi} _l\delta _{\rm RIS}^{2}}$, and reformulate \eqref{swipt_Problem1-b} as
\begin{align}\label{rewrite1}
\frac{\eta p_k\mathcal{D}_{k,l}}{\!\!\!\sum\limits_{j\ne k,k\in \mathcal{K}}{\!\!\!\eta p_j\mathcal{D}_{k,l}} \!+ \!\beta _{{\rm r},k}\pmb{\phi} _{l}^{\rm H}\pmb{\phi} _l\delta _{\rm RIS}^{2} \!+ \!\delta _{\rm r}^{2}} \!\geqslant \!\gamma _k,\, l\in {\mathcal{L}}.
\end{align}
%By doing some basic transformations, we can get 
%\setlength\abovedisplayskip{1pt}
%\setlength\belowdisplayskip{1pt}
%\begin{equation}
%	\begin{split}
%\eta &p_k\left( \pmb{\phi} _{l}^{H}\mathbf{A}_{k,l}\phi _l+2\mathrm{Re}\left\{ \mathbf{a}_{k,l}^{H}\pmb{\phi} _l \right\} +\zeta _1 \right) \\
%\geqslant& \gamma _k\sum_{j\ne k,j\in \mathcal{K} _{\varepsilon}}{\eta p_j\left( \pmb{\phi} _{l}^{H}\mathbf{A}_{k,l}\pmb{\phi} _l+2\mathrm{Re}\left\{ \mathbf{a}_{k,l}^{H}\pmb{\phi} _l \right\} +\zeta _1 \right)}\\
%&+\gamma _k\beta _{r,k}\pmb{\phi} _{l}^{H}\pmb{\phi} _l\delta _{RIS}^{2}+\gamma _k\delta _{r}^{2}.
%\end{split}
%\end{equation}
Then, we can rewrite \eqref{rewrite1} as 
\begin{align}\label{sinr1}
\pmb{\phi} _{l}^{\rm H}\mathbf{F}_{k,l}\pmb{\phi} _l+2\mathrm{Re}\left\{ \mathbf{f}_{k,l}^{\rm H} \pmb{\phi} _l \right\} \geqslant \varGamma_{k,l},\, l\in {\mathcal{L}},
\end{align}
where
\begin{align}
\mathbf{F}_{k,l}&= \eta \big( p_k-\gamma _k\sum_{j\ne k,j\in \mathcal{K} }{p_j} \big) \mathbf{A}_{k,l}-\gamma _k\beta _{{\rm r},k}\delta _{\rm RIS}^{2}\mathbf{I}_M , \\
\mathbf{f}_{k,l}&=\eta \big( p_k-\gamma _k\sum_{j\ne k,j\in \mathcal{K} }{p_j} \big) \mathbf{a}_{k,l},\\
\varGamma_{k,l} &=\gamma _k\delta _{\rm r}^{2}-\eta \big( p_k-\gamma _k\sum_{j\ne k,j\in \mathcal{K} }{p_j} \big) \beta _{{\rm d},k,l}.
\end{align}

Since we can obtain a lower bound of a convex function by using its first-order Taylor expansion, the convex lower bound of ${\pmb{\phi} _{l}^{\rm H}\mathbf{F}_{k,l}\pmb{\phi} _l}$ is expressed as
\begin{align}
	\pmb{\phi}_l ^{\rm H}\mathbf{F}_{k,l}\pmb{\phi}_l&\geqslant ({\pmb{\phi}_l}^{n})^{\rm H}\mathbf{F}_{k,l}{\pmb{\phi}_l^{n}}+2\mathrm{Re}\{ ({\pmb{\phi}_l}^{n})^{\rm H}\mathbf{F}_{k,l} \} ( \pmb{\phi}_l -{\pmb{\phi}_l^{n}} ) \notag\\
	&=2\mathrm{Re}\{ ({\pmb{\phi}_l}^{n})^{\rm H}\mathbf{F}_{k,l}\pmb{\phi}_l \} -({\pmb{\phi}_l}^{n})^{\rm H}\mathbf{F}_{k,l}{\pmb{\phi}_l^{n}},
\end{align}
where ${{\pmb{\phi}_l^{ n }}}$ is the value of ${\pmb{\phi}_l}$ at the ${n}$-th iteration.

Finally, constraint \eqref{sinr1} can be reformulated as 
\begin{align}\label{sinr2}
\!\!\!\!\!\!2\mathrm{Re}\{({\pmb{\phi}}_{l}^{n})^{\rm H}\!\mathbf{F}_{k,l}\pmb{\phi} _l\} \!\!-\!({\pmb{\phi}}_{l}^{n})^{\rm H}\!\mathbf{F}_{k,l}{\pmb{\phi}_l^{n}}\!\!\!+\!\!2\mathrm{Re}\!\left\{ \!\mathbf{f}_{k,l}^{\rm H}\pmb{\phi} _l \! \right\}\!\!\! \geqslant \!\!\varGamma_{k,l},\, \!\! l \!\in \!{\mathcal{L}}.\!\!\!\!\!
\end{align}

For non-convex constraint \eqref{SubProblem1-1-b}, we similarly derive a lower bound of ${\pmb{\phi} ^{\rm H}\hat{\mathbf{A}}_k\pmb{\phi}}$, and reformulate \eqref{SubProblem1-1-b} as
\begin{align}
		h_k\left( \pmb{\phi} \right) &\geqslant 2\mathrm{Re}\{ ({\pmb{\phi}}^{n})^{\rm H}\!\hat{\mathbf{A}}_k\pmb{\phi} \}\! -\! ({\pmb{\phi}}^{n})^{\rm H}\!\hat{\mathbf{A}}_k{\pmb{\phi}^{n}}\!\!\!+\!2\mathrm{Re}\{ \hat{\mathbf{a}}_{k}^{\rm H}\pmb{\phi} \} \!\!+\!\!{\zeta}_{1, k} \notag\\
	\label{energy2} &=2\mathrm{Re}\{ \mathbf{b}_{k}^{\rm H}\pmb{\phi} \} +\zeta _{2, k}\triangleq f_k(\pmb{\phi})\geqslant \varepsilon,
	\end{align}
where  ${\mathbf{b}_{k}^{\rm H} \!=({\pmb{\phi}}^{n})^{\rm H}\hat{\mathbf{A}}_k+\hat{\mathbf{a}}_{k}^{\rm H}}$ and ${\zeta _{2, k}={\zeta}_{1, k}-({\pmb{\phi}}^{n})^{\rm H}\hat{\mathbf{A}}_k{\pmb{\phi}^{ n }}}$.

Then, by replacing \eqref{swipt_Problem1-b}, \eqref{swipt_Problem1-d} and \eqref{energy4} with \eqref{sinr2}, \eqref{ris-energy} and \eqref{energy2}, respectively, Problem \eqref{SubProblem1} is reformulated as
	\begin{subequations}\label{SubProblem1-2}
	\begin{align}
		\max_{{\pmb{\phi}}, \varepsilon} &\quad  \varepsilon
		\\
		\mathrm{s}.\mathrm{t}.&\quad f_k(\pmb{\phi}) \geqslant \varepsilon, k\in\mathcal{K},
		\\
		&\quad \eqref{ris-energy},\eqref{sinr2},\notag
	\end{align}
\end{subequations}
which can be solved using CVX.
By solving Problem \eqref{SubProblem1-2} at each SCA iteration, the generated solutions are guaranteed to converge to the Karush-Kuhn-Tucker solution of Problem \eqref{SubProblem1} \cite{14}.
\vspace{-1ex}
\subsection{The Trajectory Design of UAV}
\vspace{-1ex}
From Problem \eqref{swipt_Problem1}, the subproblem for ${\mathbf{q}_{\rm V}}$ and ${\mathbf{t}}$ with given ${\{\pmb{\phi}_l\}}$ can be formulated as
\begin{subequations}\label{Trajectory_Problem}
	\begin{align}
		\min_{{\mathbf{q}_{\rm V}},\mathbf{t}} \label{fuction2}&\quad  E_{\rm V}\left( \mathbf{q}_{\rm V}, \mathbf{t}\right)
		\\
		\mathrm{s}.\mathrm{t}.\label{fort} &\quad t_l \geqslant 0,\, l \in \mathcal{L},
		\\
		&\quad \eqref{swipt_Problem1-b},\,\eqref{swipt_Problem1-c},\,\eqref{ris-energy}.\notag
	\end{align}
\end{subequations}

As constraints \eqref{swipt_Problem1-c} and \eqref{ris-energy} are non-convex for this subproblem, Problem \eqref{Trajectory_Problem} cannot be solved by CVX directly. Due to the strong coupling between ${\tilde{P}_{k,l}}$ and ${\mathbf{q}_{\rm V}}$ in \eqref{energy1}, \eqref{swipt_Problem1-c} is difficult to handle. Since the value of ${\mathbf{q}_{\rm V}}$ generally changes little in each iteration, by denoting ${\pmb{\psi} _{k,l}^{n}}$ as the value of ${\pmb{\psi} _{k,l}}$ at the ${n}$-th iteration, $\tilde{P}_{k,l}$ can be approximated as
\begin{align}\label{energy3}
\tilde{P}_{k,l}\approx &\left( 1-\eta \right) p_k( ( U_{1,k,l}^{n}+U_{3,k,l} ) \beta _{{\rm t},l}\notag\\
&+U_{2,k,l}^{n}\sqrt{\beta _{{\rm d},k,l}\beta _{{\rm t},l}}+\beta _{{\rm d},k,l} ), 
\end{align}
where
\begin{align}
& U_{1,k,l}^{n}=\frac{\mu _{\rm r}\mu _{\rm t}\beta _{{\rm r},k}}{\left( \mu _{\rm r}+1 \right) \left( \mu _{\rm t}+1 \right)}\pmb{\phi} _{l}^{\rm H}\pmb{\psi} _{k,l}^{n}(\pmb{\psi} _{k,l}^{n})^{\rm H}\pmb{\phi} _l,\\ 
& U_{2,k,l}^{n}=2\sqrt{\frac{\mu _{\rm d}\mu _{\rm r}\mu _{\rm t}\beta _{{\rm r},k}}{\left( \mu _{\rm d}\!+1 \right) \left( \mu _{\rm r}\!+1 \right) \left( \mu _{\rm t}\!+1 \right)}}\!\mathrm{Re}\!\left\{ \!(\pmb{\psi} _{k,l}^{n})^{\rm H}\pmb{\phi} _l \! \right\}\!\!, \\
& U_{3,k,l}=\frac{\left( \mu _{\rm r}+\mu _{\rm t}+1 \right) \beta _{{\rm r},k}}{\left( \mu _{\rm r}+1 \right) \left( \mu _{\rm t}+1 \right)}\pmb{\phi} _{l}^{\rm H}\pmb{\phi} _l. 
\end{align}
Hence, we approximate \eqref{swipt_Problem1-c} as
\begin{align}\label{function1}
 &\frac{\left( 1-\eta \right)}{E_{k}^{\rm req}}\sum_{l\in {\mathcal{L}}}p_k( ( U_{1,k,l}^{n}+U_{3,k,l} ) \beta _{{\rm t},l} \notag\\
&+U_{2,k,l}^{n}\sqrt{\beta _{{\rm d},k,l}\beta _{{\rm t},l}}+\beta _{{\rm d},k,l} ) t_l\geqslant 1.
\end{align}
In order to transform \eqref{function1} into a convex constraint, we derive the lower bounds of ${\beta _{{\rm d},k,l}}$ and ${\beta _{{\rm t},l}}$  with its first-order Taylor expansion as follows
\begin{align}
\!\!\!\!\!\!\beta _{{\rm d},k,l}\geqslant &\frac{\beta _0}{( | q_{{\rm V},l}^{n}-q_{{\rm S},k} |^2+H_{\rm V}^{2} ) ^{\frac{\tau _{\rm d}}{2}}}\notag\\
&-\frac{\tau _{\rm d}\beta _0( | q_{{\rm V},l}-q_{{\rm S},k} |^2-| q_{{\rm V},l}^{n}-q_{{\rm S},k} |^2 )}{2( | q_{{\rm V},l}^{n}-q_{{\rm S},k} |^2+H_{\rm V}^{2} ) ^{\frac{\tau _{\rm d}}{2}+1}}\triangleq \bar{\beta}_{{\rm d},k,l},\!\!\\
\beta _{{\rm t},l}\geqslant &\frac{\beta _0}{( | q_{{\rm V},l}^{n}-q_{\rm R} |^2+( H_{\rm V}-H_{\rm R} ) ^2 ) ^{\frac{\tau _{\rm d}}{2}}}\notag\\
&-\frac{\tau _{\rm t}\beta _0( | q_{{\rm V},l}-q_{\rm R} |^2-| q_{{\rm V},l}^{n}-q_{\rm R} |^2 )}{2( | q_{{\rm V},l}^{n}-q_{\rm R} |^2+( H_{\rm V}-H_{\rm R} ) ^2 ) ^{\frac{\tau _{\rm t}}{2}+1}}\triangleq \bar{\beta}_{{\rm t},l},
\end{align}
where ${q^{n}_{{\rm V}, l}}$ is the value of ${q_{{\rm V}, l}}$ at the ${n}$-th iteration. By setting the slack variables ${\{{\tilde z}_{1, l}\}}$, ${\{{\tilde z}_{2, k, l}\}}$, ${\{{\tilde z}_{3, k, l}\}}$ and ${\{{\tilde y}_{k, l}\}}$, constraint \eqref{function1} can be transformed as
\begin{align}\label{function2}
		W_{k, l}\geqslant \frac{E_{k}^{\rm req}{\tilde y}_{k,l}^{2}}{\left( 1-\eta \right) p_kt_l},
	\end{align}
where
\begin{align}
	&W_{k, l} = ( U_{1,k,l}^{n}+U_{3,k,l} ){\tilde z}_{1,l} +U_{2,k,l}^{n}{\tilde z}_{3,k,l}+{\tilde z}_{2,k,l},\\
\label{set2} &\bar{\beta}_{{\rm t},l}\geqslant {\tilde z}_{1,l} \geqslant 0,\, l \in \mathcal{L},\\
\label{set3} &\bar{\beta}_{{\rm d},k,l}\geqslant {\tilde z}_{2,k,l}\geqslant 0,\, l \in \mathcal{L},\, k\in\mathcal{K},\\
\label{set4} &{\tilde z}_{1,l} \geqslant (1/{\tilde z}_{2,k,l}){\tilde z}_{3,k,l}^{2},\, l \in \mathcal{L},\, k\in\mathcal{K}, \\
\label{set1} &\sum_{l\in {\mathcal{L}}}{{\tilde y}_{k,l}^{2}}\geqslant 1,\, k\in\mathcal{K}.
\end{align}
Similarly, constraint \eqref{set1} is transformed as
\begin{align}\label{set5}
\sum_{l\in {\mathcal{L}}}{( 2{\tilde y}_{k,l}^{n}{\tilde y}_{k,l}-({\tilde y}_{k,l}^{n})^2 ) \geqslant 1},\, k\in\mathcal{K},
\end{align}
where ${{\tilde y}_{k,l}^{n}}$ is the value of ${{\tilde y}_{k,l}}$ at the ${n}$-th iteration.

\begin{algorithm}[t] %算法开始
	\caption{SCA-based algorithm for solving Problem (13)} %算法的题目
	\label{SCA2} %算法的标签
	\textbf{Initialize}: the feasible ${\pmb{\phi}^{(1)}}$, ${\mathbf{q}_{\rm V}^{(1)}}$ and ${\mathbf{t}^{(1)}}$, the maximum iterations ${x_{\rm max}}$, ${n_{\rm max}}$ and ${r_{\rm max}}$, error tolerance ${\sigma}$.
	\begin{algorithmic}[1]
		\For{$x=1$ to $x_{max}$}
		\State {Set $n=1$, $r=1$};
		\While{${n \leqslant n_{\rm max}}$}
		\State {Given ${\pmb{\phi}^{(1)}}$, calculate ${\mathbf{q}_{\rm V}^{(n+1)}}$ and ${\mathbf{t}^{(n+1)}}$ by solving 
			\Statex \ \ \ \ Problem \eqref{Trajectory_Problem2}};
		\State {Set ${n+1 \to n}$}
		\EndWhile
		\State {Output ${\mathbf{q}_{\rm V}^{[1]} = \mathbf{q}_{\rm V}^{(n+1)}}$ and ${\mathbf{t}^{[1]} = \mathbf{t}^{(n+1)}}$}
		\While{$|f(\pmb{\phi}^{(r \!+1)})\!- \!\! f(\pmb{\phi}^{(r)})| > \!\sigma f(\pmb{\phi}^{(r)})$ and ${r \!\leqslant r_{\rm max}}$}
		\State {Given ${\mathbf{t}^{[1]}}$ and ${\mathbf{q}_{\rm V}^{[1]}}$, calculate ${\pmb{\phi}^{(r+1)}}$ by solving 
			\Statex \ \ \ \ Problem \eqref{SubProblem1-2}};
		\State {Set ${r+1 \to r}$}
		\EndWhile
		\State {Output ${\pmb{\phi}^{[1]} = \pmb{\phi}^{(r+1)}}$}
		\If {${\rm min}\{f(\pmb{\phi}^{[1]})\} < {\rm min}\{f(\pmb{\phi}^{(1)})\}$}
		\State {Set $\pmb{\phi}^{(1)}  \to \pmb{\phi}^{[1]}$};
		\Else \ \ {Set ${\pmb{\phi}^{[1]} \to \pmb{\phi}^{(1)}}$};
		\EndIf
		\EndFor
	\end{algorithmic}
\end{algorithm}

To address the non-convex constraint \eqref{ris-energy}, we first derive a lower bound of ${\beta_{{\rm t}, l}}$. 
By defining $\mathcal{Q}_l \triangleq | q_{{\rm V},l}- q_{\rm R} |^2=( q_{{\rm V},l}^{x}-q_{{\rm R}}^{x} ) ^2 +( q_{{\rm V},l}^{y}-q_{{\rm R}}^{y} ) ^2$, and utilizing the first-order Taylor expansions to ${q_{{\rm V},l}^{x}}$ and ${q_{{\rm V},l}^{y}}$, we have
\begin{align}
		\mathcal{Q}_l\geqslant & ( (q_{{\rm V},l}^{x})^n-q_{{\rm R}}^{x} ) ^2\!\!+( (q_{{\rm V},l}^y)^n-q_{{\rm R}}^{y} ) ^2 \notag\\
		&+2(  (q_{{\rm V},l}^x)^n-q_{{\rm R}}^{x} ) ( q_{{\rm V},l}^{x}- (q_{{\rm V},l}^x)^n ) \notag\\
		& +2( (q_{{\rm V},l}^y)^n-q_{{\rm R}}^{y} ) ( q_{{\rm V},l}^{y}-(q_{{\rm V},l}^y)^n ) \triangleq \bar{\mathcal{Q}_l},
	\end{align}
where ${(q_{{\rm V},l}^x)^n}$ and ${(q_{{\rm V},l}^y)^n}$ are the values of ${q_{{\rm V},l}^{x}}$ and ${q_{{\rm V},l}^{y}}$ at the ${n}$-th iteration, respectively.

Then, we can reformulate constraint \eqref{ris-energy} as 
\begin{align}\label{ris-energy2}
	\!\!\!\!\big( \sum_{k \in \mathcal{K}}{p_k}\frac{\beta _0}{( \bar{\mathcal{Q}_l} \!+\!\left( H_{\rm V}\!-\! H_{\rm R} \right) ^2 ) ^{\frac{\tau _{\rm t}}{2}}}+\delta _{\rm RIS}^{2} \big) \pmb{\phi} _{l}^{\rm H}\pmb{\phi} _l t_l \leqslant E^{\rm act}_{\rm RIS}.
\end{align}
By utilizing the first-order Taylor expansion with respect to ${\{t_l\}}$, constraint \eqref{ris-energy2} can be reformulated as
\begin{align}\label{ris-energy3}
	\big( \sum_{k \in \mathcal{K}}{p_k}\frac{\beta _0}{( \bar{\mathcal{Q}_l} \!+\!\left( H_{\rm V}\!-\! H_{\rm R} \right) ^2 ) ^{\frac{\tau _{\rm t}}{2}}}+\delta _{\rm RIS}^{2} \big) \pmb{\phi} _{l}^{\rm H}\pmb{\phi} _l \notag\\
	\leqslant E^{\rm act}_{\rm RIS}(\frac{1}{t_l^{n}}-\frac{1}{(t^{n}_l)^2}( t_l-t^{n}_l ) ),\, l\in \mathcal{L},
\end{align}
where ${t_l^{n}}$ is the value of ${t_l}$ at the ${n}$-th iteration.

For constraint \eqref{swipt_Problem1-b}, we reformulate it into
\begin{align}\label{rewrite2}
	\frac{\eta p_kW_{k, l}}{\sum\limits_{j\ne k, j \in \mathcal{K}}{\eta p_jW_{k, l}}+\beta _{{\rm r},k}\delta _{\rm RIS}^{2}\pmb{\phi} _{l}^{\rm H}\pmb{\phi} _l+\delta _{\rm r}^{2}}\geqslant \gamma _k.
\end{align}
With some basic transformations, we can rewrite \eqref{rewrite2} as
\begin{align}\label{sinr3}
	&\eta \big( \! p_k-\gamma _k\sum\limits_{j\ne k, j \in \mathcal{K}}{p_j} \!\big) \! W_{k, l} \!\geqslant \!\gamma _k( \beta _{{\rm r},k}\delta _{\rm RIS}^{2}\pmb{\phi} _{l}^{\rm H}\pmb{\phi} _l+\delta _{\rm r}^{2} ). 
\end{align}

Finally, by defining ${\Upsilon}$ as the set of slack variables ${\{{\tilde z}_{1,l}\}}$, ${\{{\tilde z}_{2,k,l}\}}$, ${\{{\tilde z}_{3,k,l}\}}$ and ${\{{\tilde y}_{k,l}\}}$, we obtain the problem to be solved at the $n$-th SCA iteration for Problem \eqref{Trajectory_Problem} as follows
\begin{subequations}\label{Trajectory_Problem2}
\begin{align}
	\min_{{\mathbf{q}_{\rm V}},\mathbf{t},\Upsilon} &\quad  E_{\rm V}\left( \mathbf{q}_{\rm V}, \mathbf{t}\right)
	\\
	\mathrm{s}.\mathrm{t}.\!\!\!\!\quad&\quad \eqref{fort}, \, \eqref{function2},\eqref{set2}-\eqref{set4},\eqref{set5},\eqref{ris-energy3},\eqref{sinr3}.\notag
\end{align}
\end{subequations}
Problem \eqref{Trajectory_Problem2} is convex and thus can be solved using CVX.

\subsection{Algorithm Development}

Based on the above discussions, we propose an SCA-based algorithm for solving Problem \eqref{swipt_Problem1}, of which details are summarized in Algorithm 1. Furthermore, we update the value of ${\pmb{\psi} _{k,l}}$ when solving Problem  \eqref{Trajectory_Problem2} to limit the approximation gap in \eqref{energy3}. Since the optimal solution to Problem \eqref{Trajectory_Problem2} may not satisfy constraint \eqref{function2}, the objective function value of \eqref{swipt_Problem1} generated by Algorithm 1 may not decrease monotonically. However, the simulation results show the fluctuations are not significant.
\begin{figure}[t]
	\centering
	\begin{minipage}[t]{1\textwidth}
		\setlength{\abovecaptionskip}{-1ex} 
		\includegraphics[scale=0.5]{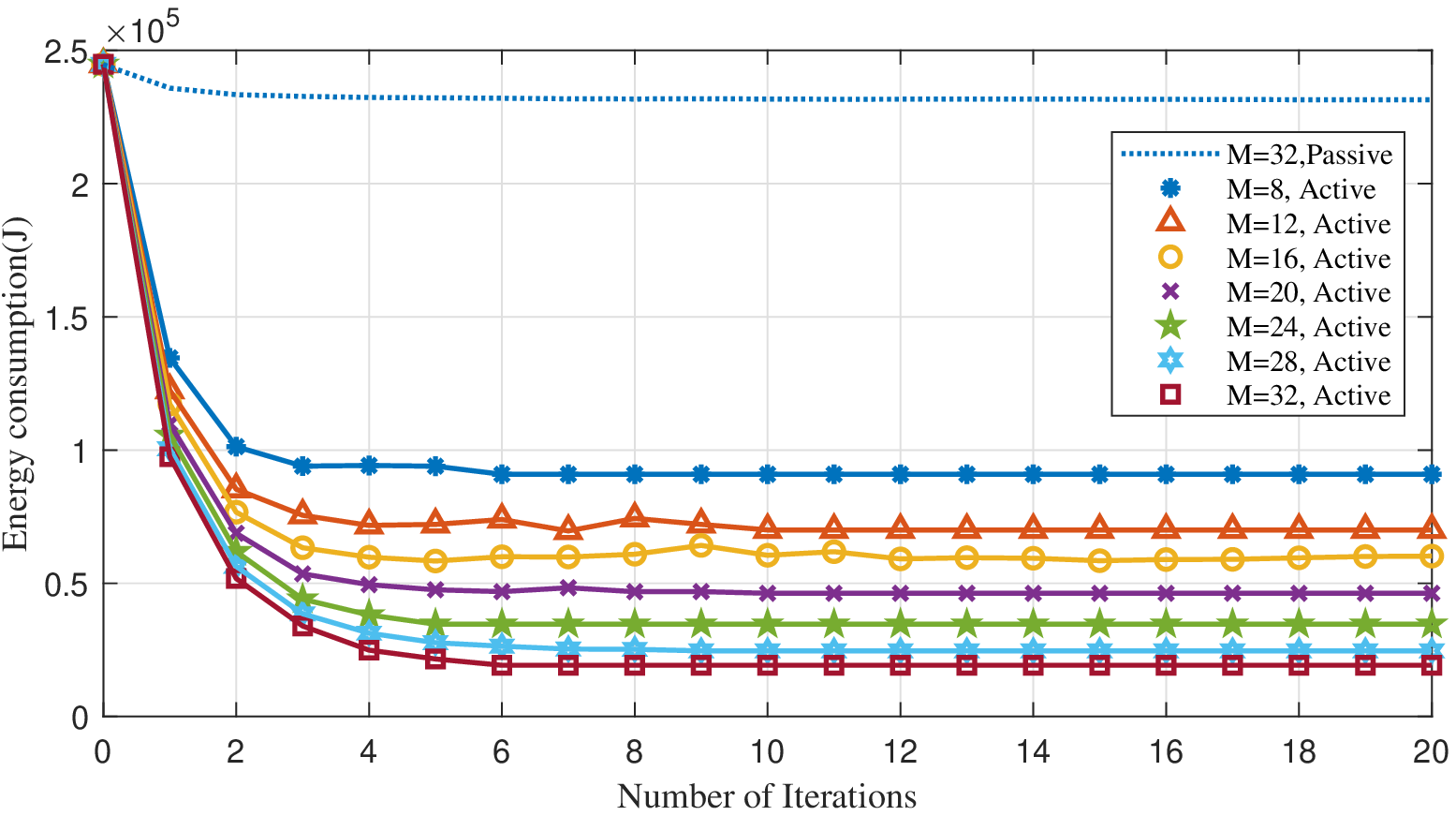}
		\caption{{The total energy cost versus the number of reflecting elements.} } 
		\label{fig2}
		
	\end{minipage}
	
	\begin{minipage}[t]{1\textwidth}
		\setlength{\abovecaptionskip}{-1ex} 
		\includegraphics[scale=0.5]{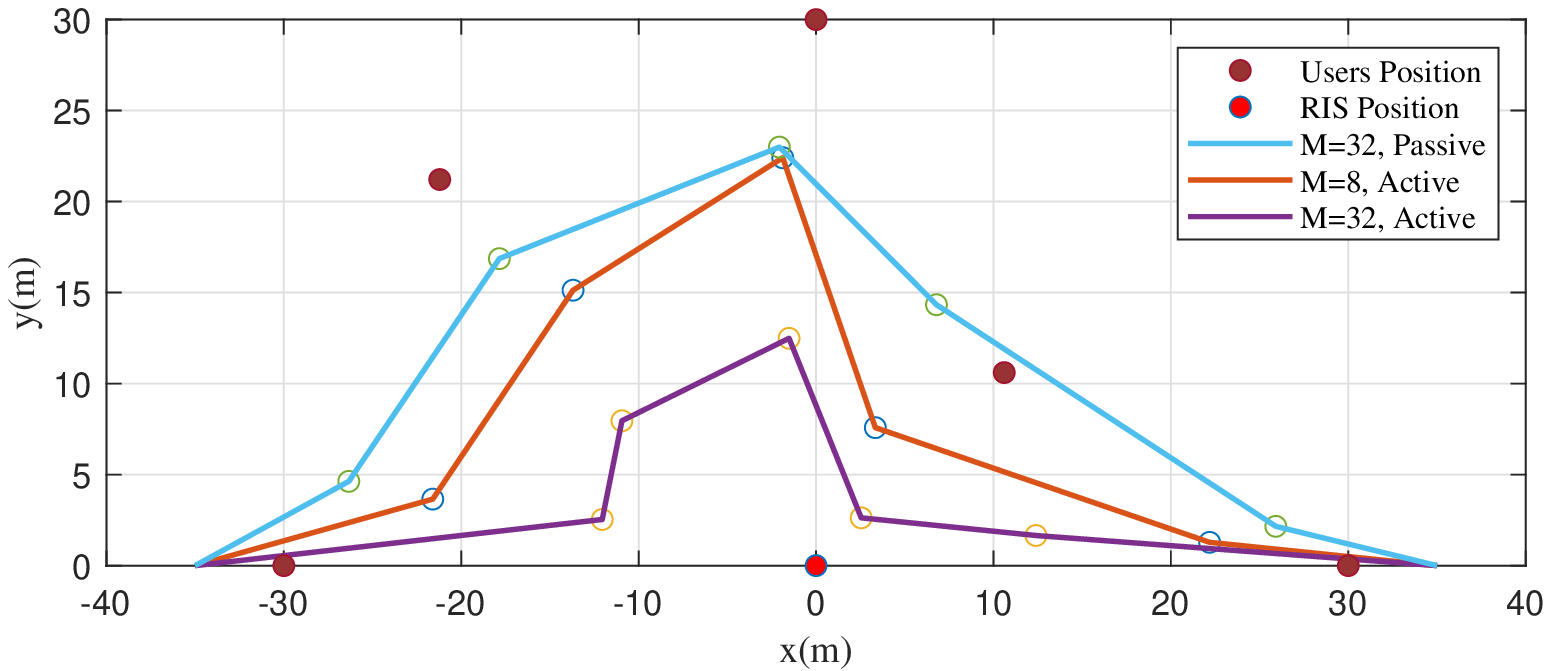}
		\caption{{The optimized trajectory of UAV under FHB protocol.} } 
		\label{fig3}
		\vspace{-1ex}
	\end{minipage}
\end{figure}

\section{Simulation Results}
To provide numerical results of our system, 5 users are assumed in a semicircular area with a radius of ${30 \,\rm m}$, and the positions of user S1-S5 are ${ (-30, 0)}$, ${ (-15\sqrt{2}, 15\sqrt{2})}$, ${(0, 30)}$, ${ (30, 0)}$ and ${(15\sqrt{2}/2, 15\sqrt{2}/2)}$, respectively, i.e., user S1-S4 are placed on the circle, and user S5 is placed at the midpoint of the radius. The position of the active RIS is ${(q_{\rm R}^x,q_{\rm R}^y)=(0,0)}$ with ${H_{\rm R} = 10\,{\rm m}}$, the initial position ${q_{{\rm V}, 0} = (-35,0)}$ and the final position ${q_{{\rm V}, L} = (35,0)}$. The UAV's flight height is set as ${H_{\rm V} = 20\, {\rm m}}$, and the transmit power ${P_{k} = 0.2 \,{\rm w}}$.
The SINR threshold of the users and the energy constraint of the active RIS are set as ${\gamma \!=\!-10\,{\rm dB}}$ and ${E_{\rm RIS}^{\rm act} \!=\!20\,{\rm J}}$, respectively. 
The parameters related to the propulsion power of the UAV are set as the same in \cite{7}, where the maximum-range speed is ${v = 18.3 \,{\rm m/s}}$. Unless stated otherwise, we set the Rician factors ${\mu_{\rm t} = \mu_{\rm r} = \mu_{\rm d} = 10}$, the path-loss exponents ${\tau_{\rm t} = \tau_{\rm r} = 2.3}$ and ${\tau_{\rm d} = 2.4}$, the energy requirement for each user ${E_{\rm req} = 0.04 \,{\rm mJ}}$, ${\eta = 0.5}$, ${\lambda = 1 \,{\rm m}}$, and ${d = 0.5 \,{\rm m}}$.

To compare the performance of two types RIS-aided model, the energy resources of UAV are assumed to be equal in both the active RIS-aided and the passive RIS-aided systems. Specifically, the total energy cost models are expressed as 
\begin{align}\label{act}
	E_{\rm pas}&=E^{\rm pas}_{\rm V},\\
	E_{\rm act} &= E_{\rm V}^{\rm act} + (L-1)E_{\rm RIS}^{\rm act}   .
\end{align}

Fig. \ref{fig2} shows the total energy cost versus the number of reflecting elements. It is seen from the figure that the total energy cost decreases with the increase of the number of reflecting elements, which verifies the effectiveness of the RIS-aided system in energy saving. From the passive RIS-aided scheme with ${M=32}$, the active RIS performs better in terms of energy saving under the same energy resources in the UAV. Moreover, the convergence of proposed algorithm, indicates that it converges within a few iterations and exhibits no noticeable fluctuations.

Fig. \ref{fig3} shows the optimized trajectory of both the active and passive RIS-aided systems when ${M = 32}$. Note that the trajectory of the UAV in the active RIS-aided system is closer to the RIS than in the passive RIS-aided system, which shows a decreased flight range, leading to a corresponding decrease in the duration available for signal transmission.

\vspace{-1ex}
\section{Conclusions}
In this paper, an active RIS-aided UAV-enabled SWIPT system was considered. We aimed to minimize the total energy cost of the UAV by alternately optimizing the trajectories, the hovering time and the reflection vectors at the active RIS based on the SCA method. The simulation results showed that the active RIS-aided system performs better in energy saving than the passive RIS-aided system under the same energy resources in the UAV-enabled systems.

	\bibliographystyle{IEEEtran}
	\bibliography{IEEEabrv,Refer}
	
\end{document}